\documentstyle[12pt]{article}
\begin{document}
\hsize = 6.0 in
\hoffset= -0.25 in
\voffset=-.5 in
\baselineskip=15pt
\newcommand{\eqletter}{ \hfill (\theequation\alph{letter})}
\newcommand{\gm}{{(\Box+e^2\rho^2)}}
\newcommand{\eql}{\nonumber &\eqletter \cr
                  \addtocounter{letter}{1}}
\newcommand{\be}{\begin{equation}}
\newcommand{\ee}{\end{equation}}
\newcommand{\bea}{\begin{eqnarray}}
\newcommand{\eea}{\end{eqnarray}}
\newcommand{\beal}{\setcounter{letter}{1} \begin{eqnarray}}
\newcommand{\eeal}{\addtocounter{equation}{1} \end{eqnarray}}
\newcommand{\none}{\nonumber \\}
\newcommand{\Vbar}{\overline{V}}
\newcommand{\Rbar}{\overline{R}}
\newcommand{\gbar}{\overline{g}}
\newcommand{\phibar}{\overline{\phi}}
\newcommand{\fcal}{{\cal F}}
\newcommand{\gcal}{{\cal G}}
\newcommand{\gtilde}{\tilde{\cal G}}
\newcommand{\pihat}{\hat{\Pi}}
\newcommand{\req}[1]{Eq.(\ref{#1})}
\newcommand{\reqs}[1]{Eqs.(\ref{#1})}
\newcommand{\larrow}{\,\,\,\,\hbox to 30pt{\rightarrowfill}
\,\,\,\,}
\newcommand{\slarrow}{\,\,\,\hbox to 20pt{\rightarrowfill}
\,\,\,}
\newcommand{\half}{{1\over2}}
\newcommand{\Fstar}{\tilde{F}}
\begin{center}
{\bf
Exact Dirac Quantization of All 2-D Dilaton Gravity Theories}
\\[20pt]
{\sl by}\\[20pt]
{D. Louis-Martinez${}^1$, J. Gegenberg${}^2$ and  G.
Kunstatter${}^1$}\\[5pt]
{\sl
${}^1$ Dept. of Physics and Winnipeg Institute of
Theoretical Physics\\
University of Winnipeg, Winnipeg, Manitoba\\
Canada R3B 2E9
}\\[5pt]
{\sl ${}^2$ Dept. of Mathematics and Statistics\\
University of New Brunswick\\
Fredericton, N.B.\\
Canada E3B 5A3}
\end{center}
\vspace*{1.5cm}
\par
\noindent
{\bf ABSTRACT:}
The most general dilaton gravity theory in 2 spacetime dimensions
is considered. A Hamiltonian analysis is performed and the reduced
phase space, which is two dimensional, is explicitly constructed
in a suitable parametrization of the fields. The theory is then
quantized via the Dirac method in a functional Schrodinger
representation. The quantum constraints are solved exactly to yield
the (spatial) diffeomorphism invariant physical wave
functionals for all theories considered. These wave functionals
depend
explicitly on the single configuration space coordinate as well
as on the imbedding of space into spacetime (i.e. on the choice of
time).
 \\[20pt]
WIN-93-06\\
{\it September, 1993}
\clearpage
\section{Introduction}
Two dimensional dilaton gravity has recently been the subject of
much work, in part because of its relation to string theory in non-
critical dimensions\cite{strings}, and in part because of its
potential usefulness as a theoretical laboratory for examining
questions concerning the end-point of Hawking radiation\cite{cghs}.
Quantization of a variety of such models has been explicitly
carried out using a variety of
techniques\cite{henneaux,bm,nna,mikovic,thiemann}. The purpose of
this paper is to carry out the Dirac quantization of the most
general 2-D dilaton gravity theory in the functional Schrodinger
representation. By choosing a suitable parametrization for the
fields we solve the quantum constraints exactly to find the unique
quantum wave functional for all such theories. The particular
technique that we use was first applied by Henneaux\cite{henneaux}
to Jackiw-Teitelboim gravity\cite{jt}, and more recently in
Ref.\cite{gk} to
spherically symmetric gravity. Similar methods have also been used
in Ref.\cite{hori} to quantize the CGHS model coupled to
conformally
invariant matter fields.
\par
The most general action functional depending on the metric tensor
$g_{\mu\nu}$ and a scalar field $\phi$ in two spacetime dimensions,
such that it contains at most second derivatives of the fields can
be written\cite{bm}:
\be
S[g,\phi]=\int d^2x \sqrt{-g} \left( \half g^{\alpha\beta}
\partial_\alpha \phi \partial_\beta \phi - V(\phi) + D(\phi)
R\right)
\label{eq: action 1}
\ee
where $R$ is the Ricci scalar\footnote{We use the conventions for
the curvature of
Misner Thorne and Wheeler\cite{mtw}.} associated with
$g_{\mu\nu}$ and
 $V(\phi)$ is an arbitrary function of $\phi$. $D(\phi)$ must be
a differentiable function of $\phi$, such that
$D(\phi)\neq 0$, and $d D(\phi)/d\phi \neq 0$ for any
admissable
value of $\phi$. Note that if $D(\phi)=constant$, the curvature
term decouples and becomes a total divergence, leaving a single,
propagating scalar degree of freedom.
\par
As discussed in Ref.\cite{bm}, the action in \req{eq: action 1}
contains only one independent function of $\phi$. This can be
readily seen as follows: The kinetic term for the scalar field can
be eliminated by means of the
following field redefinition of the metric:
\be
\gbar_{\mu\nu} = \Omega^2(\phi) g_{\mu\nu}
\label{eq: gbar defn}
\ee
where
$\Omega(\phi)$ is the  solution to the differential
equation:
\be
\half - 2 {d D\over d \phi} {d \ln\Omega\over
d\phi} = 0
\ee
\par
In terms of $\gbar$ and a new scalar field $\phibar:= D(\phi)$ the
action takes the simple form:
\be
S=\int d^2 x \sqrt{-\gbar} (\phibar\, \Rbar -\Vbar(\phibar))
\label{eq: action 2}
\ee
where the new potential $\Vbar$ is defined by:
\be
\Vbar(\phibar):= {V(\phi(\phibar))\over \Omega^2(\phi(\phibar))}
\label{eq: Vbar defn}
\ee
Note that the conditions on $D(\phi)$ given above are sufficient,
formally to guarantee the existence of $\Vbar$. For example, in
Jackiw-Teitelboim gravity\cite{jt}, $\Vbar\propto\phibar$, while
for
the
CGHS\cite{cghs} model
$\Vbar=constant$. In the case of spherically symmetric
gravity\cite{spherical,gk} $\Vbar\propto 1/\sqrt{\phibar}$.
\par
We first outline the Hamiltonian and reduced phase space analysis
of the Lagrangian in \req{eq: action 2}\footnote{We henceforth drop
the bars over
the fields.}. The 2-D spacetime is  assumed to be locally a direct
product $R\times \Sigma$, where the spatial manifold, $\Sigma$,
can at this stage be either open or closed. The metric can be
parametrized as follows\cite{torre}:
\be
ds^2 = e^{2\rho} \left[-\sigma^2 dt^2 + (dx+Mdt)^2
   \right]
\label{eq: metric parametrization}
\ee
where $x$ is a local coordinate for the spatial manifold $\Sigma$.
In terms of this parametrization the action is (up to surface
terms):
\bea
S&=&\int dt dx\left[ {\dot{\phi}\over\sigma} (2M\rho' + 2 M'-
2\dot{\rho}) + {\phi'\over \sigma} (2\sigma\sigma'-2MM' +
2M\dot{\rho}\right.
\none
& & +\left. 2\sigma^2\rho'-2M^2\rho')- \sigma^2e^{2\rho}  V(\phi)
\right]
\label{eq: action 3}
\eea
In the
above, dots and primes denote
differentiation  with respect to time and space, respectively. The
canonical momenta associated with
the fields $\{\phi, \rho, \sigma, M\}$ are therefore:
\bea
\Pi_\phi &=& {2\over\sigma} (M\rho' +M' - \dot{\rho} )
\label{eq: pi phi}\\
\Pi_\rho&=&{2\over\sigma}(-\dot{\phi} + M\phi')
\label{eq: pi rho}\\
\Pi_\sigma &=& 0
\label{eq: pi sigma}\\
\Pi_M&=&0
\label{eq: pi M}
\eea
Clearly (\ref{eq: pi sigma})and (\ref{eq: pi M}) are primary
constraints and $\sigma$ and $M$ play the role of Lagrange
multipliers. The final canonical Hamiltonian, up to surface terms,
is:
\be
H_c = \int dx \left[ M\fcal + \sigma \gcal\right]
\label{eq: canonical hamiltonian}
\ee
where
\bea
\fcal &:=& \rho'\Pi_\rho +\phi' \Pi_\phi - \Pi'_\rho \approx 0
\label{eq: fcal}\\
\gcal &:=& 2\phi'' - 2\phi'\rho' -\half \Pi_\phi \Pi_\rho +
 e^{2\rho} V(\phi) \approx 0
\label{eq: gcal}
\eea
are secondary constraints. (The notation ``$\approx0$" denotes
``weakly vanishing in the Dirac sense\cite{dirac,hrt}.)  It
is straightforward to verify that the above constraints are
first class, and that no further constraints appear in the Dirac
algorithm. $\fcal$ is the generator of spatial
diffeomorphisms on $\Sigma$, whereas $\gcal$ is the analogue of the
Hamiltonian constraint in general relativity and generates time
translations. Moreover, $\fcal$ and $\gcal$ satisfy the standard
constraint algebra associated with diffeomorphism invariant
theories in two spacetime dimensions\cite{torre}. Note that while
$\gcal$ is quadratic in the momenta, as expected, it is
off-diagonal in field space: neither $\Pi_\phi^2$, nor $\Pi_\rho^2$
appears. This is due to the fact that, in this field
parametrization, there is no kinetic term for the scalar field.
(cf. \req{eq: action 2}). As we shall see later, this is the
crucial
feature of our chosen parametrization that allows the quantum
theory to be solved exactly using these techniques.
\par
In the non-compact case, a surface term
must be added to the canonical Hamiltonian to make Hamilton's
equations agree with the Euler Lagrange equations\cite{hrt}.
The specific form of this surface term is determined in part by the
boundary conditions on the fields, which in principle vary from one
specific model to another.
\par
For completeness we write down Hamilton's equations for the momenta
\bea
\dot{\Pi}_\phi &=& \{\Pi_\phi, H\} = -2\sigma'' - 2(\sigma\rho')'
\none
  & &\quad  - \sigma e^{2\rho}{dV\over d\phi} + (M\Pi_\phi)'\\
\dot{\Pi}_\rho&=& \{\Pi_\rho, H\} = (M\Pi_\rho)' - 2(\sigma\phi')'
-2 \sigma e^{2\rho} V(\phi)
\label{eq: hamilton's equations}
\eea
\par
Apart from the Lagrange multipliers $\sigma$ and $M$ there
are two fields, ($\{\phi,\rho\}$) and their conjugate momenta in
the canonical
Hamiltonian. Given the existence of two first class constraints,
there are
no propagating modes in the
theory. The reduced phase space is finite dimensional.
We will now show that there are two phase space degrees
of freedom.
It is useful to define
the following linear combination of the constraints:
\bea
\gtilde&:=&-e^{-2\rho}(\phi'\gcal +{1\over 2} \Pi_\rho \fcal
                  )\none
&=&\left( C[\rho,\Pi_\rho, \phi, \Pi_\phi]\right)'
\label{eq: gtilde}
\eea
where we have defined the functional $C$ by
\be
C[\rho,\Pi_\rho, \phi, \Pi_\phi] :=
 \left(e^{-2\rho}\left({\Pi_\rho^2\over4}- (\phi')^2 \right)
      - j(\phi) \right)
\label{eq: C defn}
\ee
with
$j(\phi)$ a solution to the equation:
\be
{d j(\phi) \over d\phi} = V(\phi)
\label{eq: j defn}
\ee
As long as $e^{-2\rho}\phi'\neq 0$, the set of constraints
$\{\gtilde, \fcal\}$ are equivalent to the original set
$\{\gcal, \fcal\}$.
$C$ commutes with both $\gtilde$
and $\fcal$ and \req{eq: gtilde} implies that the constant
mode of the functional $C$ is unconstrained. Thus $C$ is a
physical observable. The momentum canonically conjugate
to $C$ must also be physical in the Dirac sense, and is found to
be:
\be
P:= -\int dx {2 e^{2\rho} \Pi_\rho\over
 (\Pi_\rho^2 - 4 (\phi')^2)}
\label{eq: p defn}
\ee
It also commutes with the constraints,
and has a Poisson bracket with $C$ of:
\be
\{C, P\} = 1
\ee
As expected, both $C$ and $P$ are global variables: $C$ is
constant on $\Sigma$ while its conjugate is defined as a spatial
integral.
\par
The reduced phase space for spherically symmetric gravity has been
discussed in some detail by Thiemann and Kastrup\cite{thiemann}.
In
that case one can show\cite{gk} that for static configurations, the
variable $C$ is equal to the ADM energy of the system. The momentum
variable
$P$ on the other hand is associated with global spacetime
diffeomorphisms that transform static solutions into stationary
solutions. As discussed
in detail for 4-D cosmological models with two Killing
vectors by Ashtekar and Samuel\cite{ashtekar}, this explains an
apparent discrepancy between the size of the covariant solution
space, and the size of the reduced phase space. For example, in
spherically symmetric gravity Birkhoff's theorem implies that there
is only a one parameter family of distinct
solutions, while as argued above and in \cite{thiemann}, the
reduced phase space is two dimensional: the extra phase space
variable corresponds to spacetime diffeomorphisms that cannot be
implemented canonically. In the
present context, this can be verified by noting that an
infinitesmal change in $P$
is generated by the constraint $\gtilde$ as follows:
\bea
\delta P &=& \{ P, \int dx \xi(x) \gtilde\}\none
         &=& \int dx \xi'(x)
\label{eq: delta P 1}
\eea
This is zero in the compact case, and in the non-compact
case for test functions $\xi$ that vanish at the boundaries. Since
gauge transformations are defined precisely in terms of such test
functions, $P$ is gauge invariant in this sense and commutes with
the constraints as claimed above. However, the observable $P$ can
nonetheless be changed by a ``non-canonical" diffeomorphism; one
for
which $\xi$ does not obey the usual boundary conditions. From the
spacetime viewpoint, one must consider the following coordinate
transformation:
\be
t\to t+ f(r)
\label{eq: f(r) transformation}
\ee
For constant $f(r)$, this is merely a temporal translation.
However, if $f'(r)\neq0$ such a transformation changes a static
solution to one that is stationary. In particular, under the
infinitesmal form of \req{eq: f(r) transformation}, the lapse
function goes from zero to:
\be
\delta M= \sigma^2 f'(r)
\ee
while,
\be
\delta P = \int dx {e^{2\rho}\sigma\over \phi'} f'
\label{eq: delta P 2}
\ee
For static solutions, it can be shown that
${e^{2\rho}\sigma\over \phi'}=constant$.
A comparison of \req{eq: delta P 1} and \req{eq: delta P 2}
therefore shows
that, when starting from static solutions, $\xi(r)\propto f(r)$.
Thus, as claimed in Refs.\cite{ashtekar} and \cite{thiemann} the
observable $P$ is related to the existence of stationary but non-
static solutions to the field equations. Of course, the above
analysis is only relevant in the non-compact case: when $\Sigma$
is compact, changes in $P$ cannot be implemented by the
transformations
\req{eq: delta P 1} or \req{eq: delta P 2}.
\par
We now proceed with the Dirac quantization of the theory in the
functional Schrodinger representation, in which states are given
by functionals of the fields, namely\footnote{The constraints
$\Pi_M\approx 0$ and $\Pi_\sigma\approx 0$ will require the wave
functional to be independent of the arbitrary Lagrange multipliers
$M$ and $\sigma$. We therefore drop them from the following
discussion.}
\be
\Psi = \Psi[\phi, \rho]
\ee
As usual in the Schrodinger representation, we define the momentum
operators as (functional) derivatives:
\bea
\pihat_\phi&=& -i\hbar {\delta \quad\over \delta \phi(x)}
 \label{eq: pihat phi}\\
\pihat_\rho&=& -i\hbar {\delta \quad\over \delta \rho(x)}
\label{eq: pihat rho}
\eea
\par
The next step\cite{henneaux,gk}, is to note that the
constraints $\fcal$ and $\gcal$ (or equivalently $\fcal$ and
$\gtilde$ can be solved for the momenta as follows:
\bea
\Pi_\phi &\approx& {g[\rho, \phi]\over Q[C; \rho,\phi]}
\label{eq: Pi phi constraint}\\
\Pi_\rho&\approx& Q[C; \rho,\phi]
\label{eq: Pi rho constraint}
\eea
where we have defined:
\bea
Q[C;\phi,\rho]&:=& 2\sqrt{(\phi)'{}^2 + (C + j(\phi)) e^{2\rho}}
\label{eq: Q defn}\\
g[\phi,\rho]&:=& 4\phi'' - 4\phi'\rho' + 2e^{2\rho} V(\phi)
\label{eq: g defn}
\eea
In \req{eq: Q defn}, $C$ is a constant
of integration that corresponds precisely to the observable defined
in \req{eq: C defn}.
\par
We now define physical states $\Psi_{phys}[\phi,\rho]$, as those
that are annihilated
by the constraints:
\bea
\left(\pihat_\phi-{g[\phi,\rho] \over Q[C;
\phi,\rho]}\right)\Psi_{phys} &=& 0
\none
\left(\pihat_\rho - Q[C; \phi,\rho]\right)\Psi_{phys} &=& 0
\label{eq: quantum constraints}
\eea
The solution to these constraints take the form:
\be
\Psi_{phys}[C;\phi,\rho] = \exp \left({i\over\hbar}
S[C;\phi,\rho]\right)
\ee
where $S$ satisfies the linear, coupled functional differential
equations:
\bea
{\delta\quad\over \delta\phi} S[C;\phi,\rho] &=& {g[\phi,\rho]\over
Q[C;\phi,\rho]}\none
{\delta\quad\over\delta\rho} S[C;\phi,\rho] &=& Q[C;\phi,\rho]
\label{eq: equations for S}
\eea
The second of these equations can be directly integrated, since it
does not involve spatial derivatives, to yield:
\be
S[C;\phi,\rho] = \int dx \left[ Q + \phi' \ln\left({ 2\phi' - Q
\over
   2\phi' + Q}\right) \right] + F[C;\phi]
\label{eq: partial solution}
\ee
where $F[C;\phi]$ as an arbitrary functional independent of $\rho$.
Remarkably, the remaining operator constraint, applied to \req{eq:
partial solution} yields the simple result that $F[C;\phi]=F[C]=
constant$(independent of $\phi$). Thus
physical states are described by wave
wave functionals of the form:
\be
\Psi_{phys}[\phi,\rho] = \chi[C] \exp {i\over \hbar}
\int dx \left[ Q + \phi' \ln\left({ 2\phi' - Q \over
   2\phi' + Q}\right) \right]  ,
\label{eq: wave functional}
\ee
where $\chi[C]= \exp i/\hbar F[C]$ is a completely arbitrary
function of
the configuration space coordinate $C$. This arbitrariness in the
wave functional is a consequence of the fact that the reduced
Hamiltonian for the system (in terms of $C$ and $P$) vanishes
identically.
\par
The wave functional in \req{eq: wave functional} is invariant under
spatial diffeomorphisms
generated by the quantum constraint:
\be
\hat{\fcal} = \rho' \hat{\Pi}_\rho + \phi' \hat{\Pi_\phi}
    -{\partial\quad\over \partial x} \hat{\Pi}_\rho
\ee
It is also annihilated by the Hamiltonian constraint, with factor
ordering:
\be
\hat{G} = \half g[\phi, \rho] -\half Q \hat{\Pi}_\phi Q^{-1}
\hat{\Pi}_\rho
\ee
We also note that the phase of the wave functional becomes complex
when
$Q$ is imaginary. This is consistent with traditional quantum
mechanics: $Q^2<0$ corresponds to classically forbidden regions in
which the classical momenta are imaginary (cf. \req{eq: Pi phi
constraint}). The phase can also pick up an imaginary part when the
argument of the logarithm is negative, i.e. when
\be
4(\phi')^2-Q^2 < 0
\label{eq: imaginary condition}
\ee
The physical significance of this contribution is less clear at
this stage. However, it is interesting to note that in the case of
spherically symmetric gravity, \req{eq: imaginary condition} is
satisfied for static solutions expressed in Kruskal coordinates
when $r<2m$\cite{gk}.
\par
Finally, we remark that the wave functional $\Psi_{phys}$
yields, as expected, a consistent representation for the physical
phase space in the Schrodinger representation. That is, one can
explicitly verify the relation:
\be
-i\hbar {\partial \over \partial C}\Psi_{phys} =
   P \Psi_{phys}
\ee
where $P$ is equal to the expression in \req{eq: p defn} evaluated
on the constraint surface.
\par
The solution \req{eq: wave
functional}  has very interesting features
that highlight several important issues in quantum gravity. As the
notation indicates, the wave functional is an explicit function of
the physical configuration space coordinate $C$ as well a
functional of the imbedding variables $\rho$ and $\phi$.
Invariance of the wave functional under spatial diffeomorphisms
guarantees that one of the two functions is redundant: it can be
trivially eliminated by choosing an appropriate spatial
coordinate. The remaining function is essentially the time variable
in the problem. Different choices correspond to different time
slicings. The fact that the wave functional depends in a non-
trivial way on this choice of time slice (as opposed to the choice
of spatial coordinate) is sometimes referred to as the many-
fingered time problem in quantum gravity\cite{quantum gravity}.
However, in the present context it can also be interpreted as a
consequence of the fact that the solution \req{eq: wave functional}
which solves the Hamitonian constraint is analoguous to a time
dependent Schrodinger state.
\par
We have so far avoided the question of the correct Hilbert space
measure. One can perhaps ask whether there exists a functional
measure on
the space of $\phi$ and $\rho$ that
makes this state normalizable. Alternatively, one can first choose
a spatial coordinate and time slicing and then define the
measure only on the (one-dimensional) space of physical
observables. This question can perhaps best be studied in the
context of particular models, such as spherically symmetric
gravity or the CGHS model. In any case, it is hoped that the
existence of the above exact solution in a the most general theory
of
2-D dilaton gravity provides a useful laboratory for the study
of this, and other fundamental questions in quantum gravity.
\par
Finally we remark that in Ref.\cite{gk}, the solution \req{eq: wave
functional} was
applied to spherically symmetric gravity in order to find the
quantum wave functional for an isolated black hole. This wave
functional was shown to have interesting properties consistent with
the presence of a quantum mechanical instability
(i.e. Hawking radiation.)\footnote{Qualitatively similar results
have recently been obtained in spherically symmetric gravity by
Nakamura {\it et al}\cite{nakamura}.} There is in fact a large
class dilaton gravity theories that possess Schwarzschild type
black hole solutions. A complete  analysis of the exact quantum
wave functional for these theories,
along the lines of Ref.\cite{gk} is currently in progress.\\[20pt]
{\Large\bf Acknowledgements}
We are grateful to Steve Carlip for helpful conversations and a
critical reading of the manuscript.
This work was supported in part by the Natural Sciences and
Engineering Research Council of Canada.  \par
\end{document}